\documentclass[apl,twocolumn,superscriptaddress]{revtex4}
\usepackage[latin1]{inputenc}
\usepackage{amsmath}
\usepackage{amsfonts}
\usepackage{amssymb}
\usepackage{graphicx}
\usepackage{float} %places pictures on defined locations

\begin{document}

\title{Multiferroic heterostructures for spin filter application - an ab initio study}
\date{\today}
\author{Stephan Borek}
\author{J\"urgen Braun}
\author{Hubert Ebert}
\affiliation{Department Chemie, Ludwig-Maximilians-Universit\"{a}t M\"{u}nchen, Butenandtstra\ss e 5-13, 81377 M\"{u}nchen, Germany}
\author{J\'an Min\'ar}
\affiliation{Department Chemie, Ludwig-Maximilians-Universit\"{a}t M\"{u}nchen, Butenandtstra\ss e 5-13, 81377 M\"{u}nchen, Germany}
\affiliation{New Technologies-Research Centre, University of West Bohemia,\\ Univerzitni 8, 306 14 Pilsen, Czech Republic}
%\author{Christian Langenk\"amper}
%\author{Christian Thiede}
%\author{Markus Donath}
%\affiliation{Westf\"alische-Wilhelms Universit\"at M\"unster}
%\author{Hubert Ebert}
%\affiliation{Ludwig-Maximilaians Universit\"at M\"unchen}

\begin{abstract}
Novel imaging spin-filter techniques, which are based on low energy 
electron diffraction, are currently of high scientific interest. 
To improve the spin-detection efficiency a variety of new materials 
have been introduced in recent years. A new class of promising
spin-filter materials are represented by multiferroic systems, 
as both magnetic and electric ordering exist in these materials. 
We have investigated Fe/BaTiO3(001), which defines a prominent 
candidate due to its moderate spontaneous polarization,
for spin filter applications calculating diffraction patterns for
spin polarized electrons incident on the Fe surface. Motivated by
the fact that spin polarized low energy electron diffraction
is a powerful method for the determination of the properties
of surfaces we investigated the influence of switching
the BaTiO$_{3}$ polarization on the exchange and
spin orbit scattering as well as on reflectivity and
figure of merit.
This system obviously offers the possibility to realize
a multiferroic spin filter and manipulating the spin-orbit
and exchange scattering by an external electric field.
The calculations have been done for a large range of kinetic energies
and polar angles of the diffracted electrons considering
different numbers of Fe monolayers.
\end{abstract}

\maketitle

\section{Introduction}
Materials combining different ferroic properties are called
multiferroics \cite{schmidt1}.
Multiferroic systems, especially multiferroic heterostructures,
are promising in technical applications
giving for example the opportunity to change the magnetization 
of a ferromagnetic material altering
the electric polarization of a ferroelectric material or vice versa.
An essential requirement of the realization of such systems was
the development of experimental techniques over the last years for
growing epitaxial magnetic and ferroelectric materials \cite{dawber1,zheng1}. 
Additionally a well-defined interface has to be ensured between both
ferroic phases to enable an effective coupling.
Various experimental groups investigated prototypical
devices using external electric fields for a change of the
ferroelectric polarization and investigated the influence
on the magnetic properties of the ferromagnetic phase \cite{eerenstein1}.
The application of such new systems are broad but especially as
memory devices epitaxial thin films might be interesting \cite{ahn1}.
Because of the technical applicability the discovery for new multiferroic
materials with strong coupling between the ferromagnetic and the ferroelectric
phases has been strengthen \cite{kimura1,lawes1}.
The understanding of the coupling mechanism between both ferroic phases is
essential for the development of multiferroic heterostructures and
their technical application.
Therefore different ferroelectric materials have been studied extensively in previous
theoretical investigations where the BaTiO$_3$ (BTO) takes a prominent position due to its
moderate spontaneous polarization \cite{fechner1}.

For the construction of a multi-channel vector spin polarimeter
at BESSY II new types of material classes are tested for
their applicability as reflecting mirror for spin polarized
electrons. Beside well tested materials likewise W, Ir, and
oxygen passivated Fe other material classes should be
investigated yielding interesting properties for
spin filter purposes. Because of the high
magnetic moments for the Fe layers of Fe/BTO and 
the dependency of their magnitude on the BTO polarization
\cite{fechner2,borek1} this heterogeneous system
is an interesting candidate for
the application as spin polarizing mirror.
In principle using a multiferroic heterostructure
would give the possibility to alter the exchange
and spin-orbit scattering switching the electrical
polarization of the BTO. Thereby no additional
magnetic field is necessary for changing the
surface magnetization of the ferromagnetic material.
Especially working areas inducing an altering of
the scattered spin direction by switching the
BTO polarization are interesting for an application
as spin polarizing mirror.

In this work we studied the scattering of spin polarized electrons from the
surface of the multiferroic heterostructure Fe/BTO.
We calculated diffraction patterns visualizing
the dependency of the scattered intensity
on the polar angles and kinetic energies.
Because the polarization of the electrons were oriented
perpendicular to the scattering plane,
both exchange and spin-orbit scattering occur \cite{feder1,tamura1}.
Therefore we investigated the 
so-called exchange and spin-orbit asymmetries as
well as the effective reflectivity and the
figure of merit (FOM).
Additionally we investigated the layer dependence
of the exchange and spin-orbit scattering.
We applied SPLEED
(Spin Polarized Low Energy Electron Diffraction)
calculations to
1, 2 and 3 monolayer (ML) Fe on top of
BTO because their applicability as multiferroic system
has been mentioned and their electronic
and magnetic structure was investigated
in detail in previous works \cite{fechner2,meyerheim1}.
For all systems the crystal structure was
taken from previous investigations providing a relaxed interface and surface \cite{fechner2}.
This accounts for the fact that SPLEED is very surface sensitive
due to the low kinetic energies of the incident electrons.
For the calculation of the electronic properties and the
electron diffraction we applied the fully relativistic KKR-method
in the framework of spin-polarized density functional theory
to account for effects based on exchange and spin-orbit
interactions in one step \cite{ebert1,SPR-KKR6.3}. 

The paper is organized as follows: In Sec. \ref{sec_theory}
we describe the application of the underlying SPLEED theory
in the framework of the SPR-KKR method.
In Sec. \ref{sec_discussion} we discuss our computational
results concerning
the electron diffraction. In Sec. \ref{sec_summary}
we summarize our results.

\section{Theoretical application \label{sec_theory}}

\begin{figure}[H]
%\hspace{-1.cm}
\centering
\includegraphics[scale=0.32,angle=0]{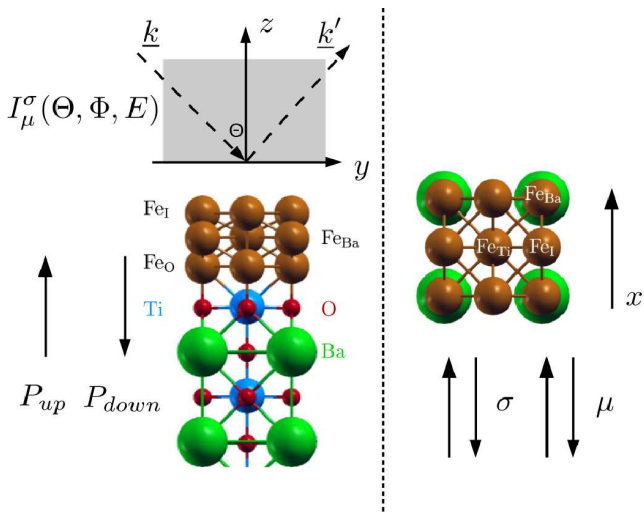}
\caption{
Half-space of 3 ML Fe/BTO used for the SPLEED calculations.
Left: Side view. Right: Top view. The various atomic
types have been indicated. The reflected electron
intensity ($I_{\mu}^{\sigma}(\Theta,\Phi,E)$) is
drawn. $\Theta$ is the polar angle of the incident
electrons and $\Phi$ the azimuthal angle.
Additionally the parameters which have been
varied are shown:
The electric polarization of BTO (P$_{up}$, P$_{down}$),
the polarization of the electron ($\sigma$),
the in plane magnetization of the surface ($\mu$).
The plane of incidence is spanned by the wave vectors
($\underline{k}, \underline{k}'$) and is aligned
along the [010]-direction. The polarization of the electron
and the surface magnetization are altered
along the [100]-direction.}
\label{scattering_principle}
\end{figure}

We briefly introduce the theoretical method implemented
in the SPR-KKR program package for the calculation
of spin-polarized electron scattering from
arbitrary surface systems.
The SPLEED calculations are done using the layered-KKR method \cite{feder2}.
This method describes the scattering of spin polarized 
electrons from a stack of atomic layers representing a
semi-infinite surface system.
The two main steps are the treatment
of multiple scattering within one specific
atomic layer and the scattering between the atomic layers
of the stack. The combination of both scattering
mechanisms result in the calculation of the
so-called bulk reflection matrix.
With the determination of the bulk reflection matrix
the diffraction of spin-polarized electrons from a surface is defined \cite{feder1,feder2}.
%For the description of the multiple scattering in an
%atomic layer the calculation of the Kambe X-matrix is required \cite{kambe1}.
%
%\begin{equation}
%X_{l'm'}=\sum_{j=1}^{\infty}G_{l'm'}t_{j}\exp(i\mathbf{k}_{||}\mathbf{R}_{j})
%\end{equation}
%
%For the scattering in between the atomic layers
%the so-called layer-doubling method is applied.
%The combination of both scattering mechanisms results in the calculation of the
%bulk-reflection matrix ($\mathbf{M}^{\pm}_{1\dots n}$). This matrix connects the
%amplitude of an incident electron ($\phi^{+}_{1,lm}$) with the amplitude of the
%scattered electron ($\phi^{-}_{1,lm}$).
%
%\begin{equation}
%	\phi^{-}_{n=1,lm}=\mathbf{M}^{\pm}_{1\dots n}\phi^{+}_{n=1,lm}
%\end{equation}
%
%The subscripts indicate the number of the different atomic layers ($n$)
%and appropriate quantum numbers like the orbital ($l$) and
%magnetic quantum number ($m$).
%With the determination of the bulk reflection matrix
%the diffraction of spin-polarized electrons from a surface is defined \cite{feder1,feder2}.
Considering exchange and spin-orbit scattering
there exist mainly two configurations according
to the orientation of magnetization and polarization of
the electron with respect to the scattering plane. 
Whereby the scattering plane is spanned
by the wave vector of the incident electron $\mathbf{k}$ and the
scattered electron $\mathbf{k}'$ (see Fig. \ref{scattering_principle}).
In our calculations both
surface magnetization and polarization of the electron are oriented
perpendicular to the scattering plane. Therefore 
exchange and spin-orbit scattering
contribute \cite{tamura1}.

The calculation of the scattered electron intensity
was done for all combinations of magnetization and
polarization, i.e. one ends up with four
different intensities ($I_{\mu}^{\sigma}$) \cite{feder1,tamura1}.
Here $\mu$ is the magnetization direction and $\sigma$
gives the polarization direction of the electron.
From the reflected intensities $I_{\mu}^{\sigma}$ the exchange asymmetry, spin-orbit
asymmetry and FOM are calculated.
The spin-orbit asymmetry can be determined via the relation \cite{feder1}

\begin{equation}
 A_{soc}=\frac{1}{2}(A_{+}-A_{-}),
\end{equation}

which can be ascribed to the reflected intensities
using the definitions

\begin{align}
 A_{+}=\frac{I_{+}^{+}-I_{+}^{-}}{I_{+}^{+}+I_{+}^{-}} \\
 A_{-}=\frac{I_{-}^{+}-I_{-}^{-}}{I_{-}^{+}+I_{-}^{-}}.
\end{align}

The exchange scattering for a specified magnetization
direction is defined as $A_{+}$ and $A_{-}$.

%Another useful quantity considered by our investigations is the asymmetry defined by \cite{tamura1}
%
%\begin{equation}
% A_{u}=\frac{I_{+}^{+}+I_{+}^{-}-I_{-}^{+}-I_{-}^{-}}{I_{+}^{+}+I_{+}^{-}+I_{-}^{+}+I_{-}^{-}}
%\end{equation}
%
%This quantity can be interpreted as asymmetry produced by an unpolarized incident electron beam upon reversal of magnetization
%direction \cite{tamura1}. It vanishes either if there is no spin-orbit coupling ($I_{\mu}^{\sigma}=I_{-\mu}^{-\sigma}$) or
%there is no exchange interaction ($I_{\mu}^{\sigma}=I_{-\mu}^{\sigma}$). In this interpretation it gives the
%strength of the dual contribution of both spin-orbit and exchange interaction.

The most important quantity for the characterization of the diffraction of electrons
from a surface is the FOM. The reflectivity
as well as the asymmetry (exchange or spin-orbit) contribute to
this quantity via the following equation

\begin{equation}
	\text{FOM}_{+(-)}=I_{+(-)}\cdot A_{+(-)}^{2}. \label{eq:fom}
\end{equation}

In Eq. (\ref{eq:fom}) the FOM is calculated separately for the two specific
magnetization directions. This gives insight into
changes of the scattering behaviour 
changing the magnetization direction at the surface.
The spin orbit induced FOM is defined by

\begin{equation}
        \text{FOM}_{soc}=I_{eff}\cdot A_{soc}^{2},
\end{equation}

with $I_{eff}$ the effective reflectivity

\begin{equation}
	I_{eff}=I_+^++I_+^-+I_-^++I_-^-.
\end{equation}

%\section{Electronic structure}
%The calculation of the electronic structure was discussed
%intensively in previous works \cite{fechner2,borek1}.

For the calculation of SPLEED patterns it is necessary
to determine the single site scattering matrix for the
individual atomic types involved in the half-space of Fe/BTO.
The self-consistent potentials necessary for calculating the
single-site scattering matrices were taken from
previous works \cite{fechner2,borek1}.
With the single site-scattering matrix we were able
to calculate the Kambe X-matrix and the bulk
reflection matrix as described above.

%According to the experimental configuration different scattering mechanism
%can occur in relation with the symmetry of the surface \cite{tamura1}.
%We used a configuration were both the polarization of the electron
%as well as the magnetization of the surface are aligned perpendicular
%to the scattering plane which is stretched by the wave vector of the
%incident and the scattered electron. In this case both spin-orbit
%and exchange scattering occurs.
%We calculated the four scattering intensities $I_\mu^\sigma=\{I_+^+,I_+^-,I_-^+,I_-^-\}$
%according the orientation of polarization ($\sigma$) of the electron and their relation to 
%the orientation of the magnetization of the Fe surface ($\mu$) .

Considering the escape of the diffracted electrons into vacuum
an essential quantity for SPLEED is the work function and the surface potential
barrier. For the work function we applied 4.7 eV which is a reasonable
value for Fe(001) \cite{skriver1} whereas
the surface potential barrier was simulated by the Rundgren-Malmstr\"om 
parametrization \cite{rundgren1}.
We calculated SPLEED patterns for a broad range of kinetic energies and polar
angles according to the working areas as scattering mirror. 
The calculations were done for the specular reflected beam
and for the polarization of the BTO pointing in the direction of the
surface normal (P$_{up}$) and to the opposite direction (P$_{down}$).

In Fig. \ref{scattering_principle} the side view and
top view of the Fe/BTO heterostructure are shown. Here
3 ML Fe are placed on top of the substrate BTO. The unit cell of BTO
corresponds to a tetragonal distorted structure (P4mm) with
a lattice parameter of 3.943 \AA \cite{fechner3}.
The structural details have been published in
previous works \cite{fechner2,borek1}.

Remarkably the surface of the BTO has
a (001) orientation and is terminated by O and Ti.
The termination of the Fe/BTO system has been
investigated in previous experimental studies \cite{MKE+11}.
The multiferroic heterostructure was simulated
by setting up a half-space of four unit cells
of BTO for the transition from the surface to the
bulk. From the fifth unit cell on the bulk
region was started. Due to the distorted unit cell
of BTO (P4mmm) a remanent
electrical polarization occurs. This result from
a shift of the Ti atom and the O atoms
in opposite direction and generates
the ferroelectric properties of BTO \cite{borek1}.
It has been shown that the first Fe layer (Fe$\rm_O$) 
on top of BTO is positioned on top of
the O atoms. The second Fe layer has
two inequivalent Fe sites which are
on top of Ti (Fe$\rm_{Ti}$) and Ba (Fe$\rm_{Ba}$). 
The third Fe layer (Fe$\rm_I$) is placed on top
of the Fe atoms of the first Fe layer \cite{fechner2}.
The surface as well as the interface of the
multiferroic heterostructure has been relaxed
in previous works \cite{fechner2}
using the Vienna Ab initio Simulation Package (VASP) \cite{KF96,KJ99}.
The resulting structure parameters have been
presented and discussed \cite{fechner2}.
Using the relaxed crystal structure we applied
the fully relativistic multiple scattering
formalism for the calculation of the SPLEED patterns
as sketched above \cite{SPR-KKR6.3}.
The self-consistent calculation of the
system potentials has been done using a previously
multicode approach \cite{borek1,LET+01,Ebert96}.

\section{Results and Discussion \label{sec_discussion}}

The results of the electronic structure calculations,
especially the spin magnetic moments
have shown to be in good agreement with
published data \cite{fechner2,borek1}.
The magnetic properties of the Fe surface
are most important for the exchange scattering
of spin-polarized electrons. Changing the
polarization of the BTO affects the
magnetic moments of the Fe layers. Therefore
the surface magnetic moments have shown
to react on a competition of the
reduced number of nearest neighbors at the surface
and hybridization effects
between the Fe $3d$ and O $2p$ states as well as
the Fe $3d$ and Ti $3d$ states \cite{DSM+06,fechner1,fechner2}.
The impact on the electronic structure has been investigated
in detailed studies \cite{fechner2}.
It was shown that for 1 ML Fe on
top of BTO a ferromagnetic ground state occurs
with a large spin magnetic moment for Fe due
to the reduced coordination number at the surface.
For 2 ML Fe on BTO a ferrimagnetic ground state was
predicted whereas the Fe atoms in the second Fe layer
(Fe$_{\rm Ti}$, Fe$_{\rm Ba}$) show an antiparallel alignment
of their spin directions and different magnitudes.
For 3 ML Fe a ferromagnetic ground state was
predicted. The change of the BTO polarization
affects the spin magnetic moments of
Fe via a change in the hybridization of
the Fe, Ti and O states mentioned above.
A smaller (larger) hybridization of the
$3d$ states of Fe - Ti and the $2p$ states
of O results in a larger (smaller) spin
magnetic moment of Fe \cite{borek1}.
Therefore the magnetoelectric coupling between the ferroelectric
and ferromagnetic material is based on
hybridization effects \cite{fechner2}.
%Because of changes in the crystal structure at the interface
%when switching the polarization of the BTO the magnetic and
%electronic properties of the Fe layers are affected \cite{fechner2}.
The change of the magnetic properties has an influence 
on the optical properties like
the absorptive part of the conductivity tensor \cite{borek1}
and are detectable using the scattering of electrons with low
kinetic energy, as will be shown below.

In Fig. \ref{theta_energy_maps_refl_p_pup_pdown_3ml} the reflectivity for 
3 ML Fe/BTO for a [100] surface magnetization is shown.

\begin{figure}[H]
%\hspace{-1.cm}
\begin{minipage}[c]{.2\textwidth}
\includegraphics[scale=0.30,angle=270,clip=true,trim=2cm 4.cm 1.cm 7.5cm]{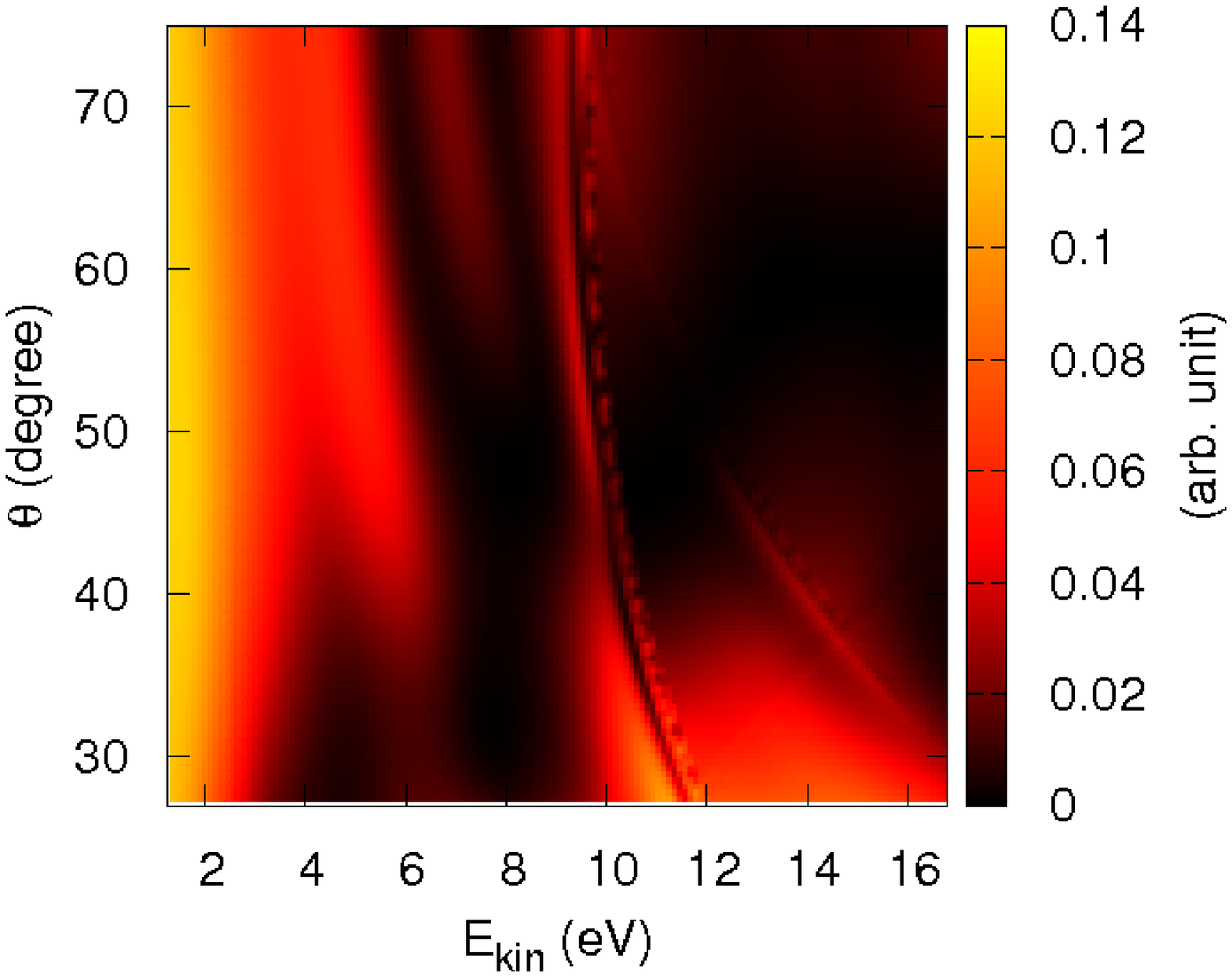}
\end{minipage}
\hspace{0.3cm}
\begin{minipage}[c]{.2\textwidth}
\includegraphics[scale=0.30,angle=270,clip=true,trim=2cm 6.2cm 1.cm 4cm]{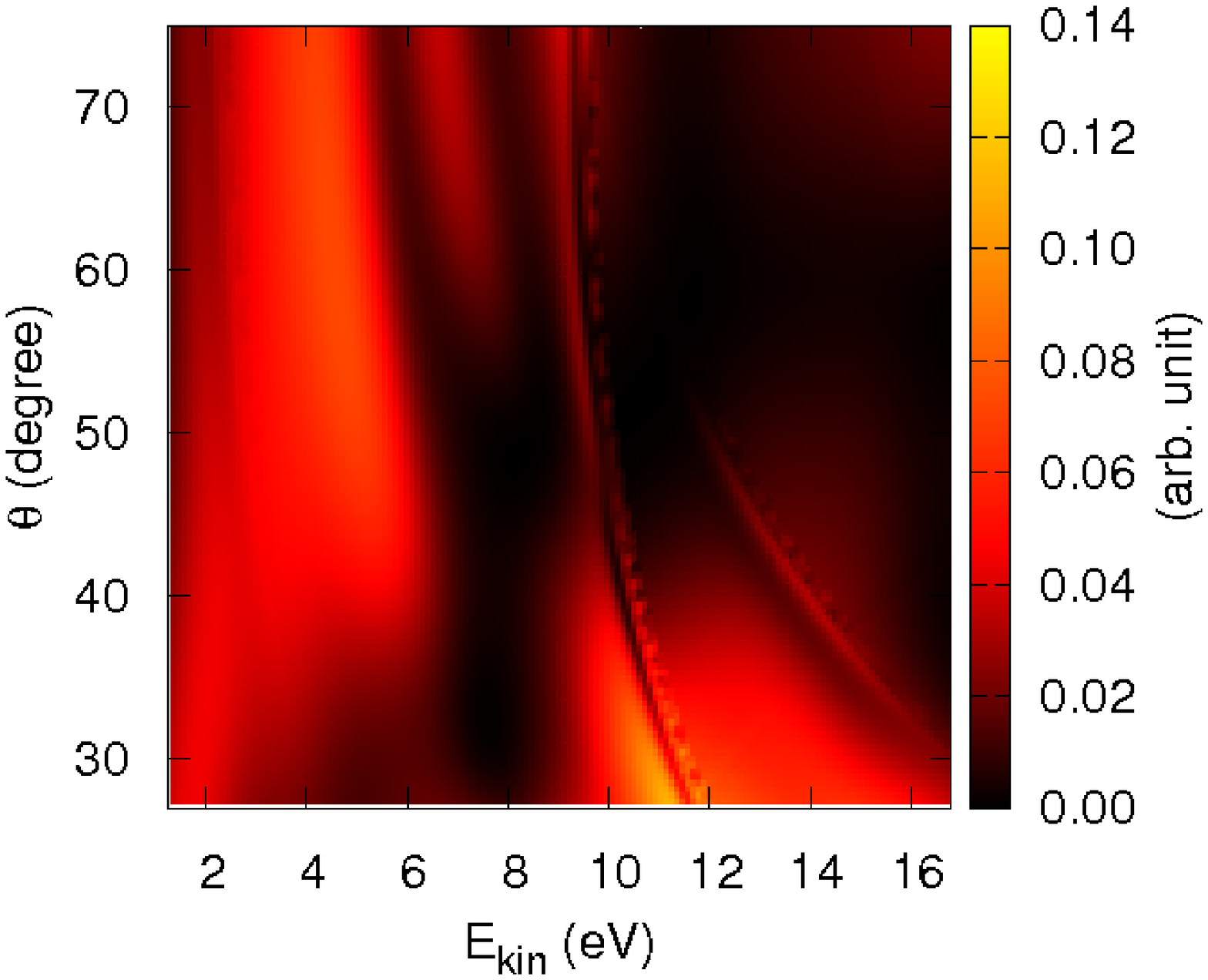}
\end{minipage}
\caption{Diffraction pattern of the reflectivity for 3 ML Fe/BTO for a [100]-orientation
of the magnetization direction (color online). 
Left: The polarization of the BTO is directed along the surface normal (P$_{up}$). 
Right: The polarization of the BTO is directed in opposite direction (P$_{down}$).}
\label{theta_energy_maps_refl_p_pup_pdown_3ml}
\end{figure}

The range of polar angles and kinetic energies have been chosen
according to a possible application as scattering mirror \cite{thiede1}.
The diffraction pattern can be divided into two main
parts with respect to the kinetic energy.
The main differences occur for kinetic energies $\le$ 8 eV,
i.e. below the emergence threshold.
The emergence threshold marks the occurrence of an additional
beam lowering the intensity of the specular diffracted one.
The area of high reflectivity for kinetic energies
at 4 eV covers nearly the total range of polar angles.
This is important for later application as scattering mirror
giving the possibility to vary the position of the
mirror with respect to the polar angle of the incident electrons.
For kinetic energies lower the emergence threshold
a large sensitivity changing the BTO polarization
is visible. On the other hand for kinetic energies above the
emergence threshold switching the BTO polarization
has less influence on the reflectivity. This
is due to the fact that electrons with lower kinetic
energy react more sensitive on changes of the
surface magnetization, i.e. onto variations of
the exchange and spin-orbit scattering.
For kinetic energies above 10 eV an area of
high reflectivity is visible for polar angles
around 30$^\circ$. 
Because of the high kinetic energy
this area is much less affected by
a change of the BTO polarization.

\begin{figure}[h]
\includegraphics[scale=0.4,clip=true,trim=0cm 6cm 0cm 2cm]{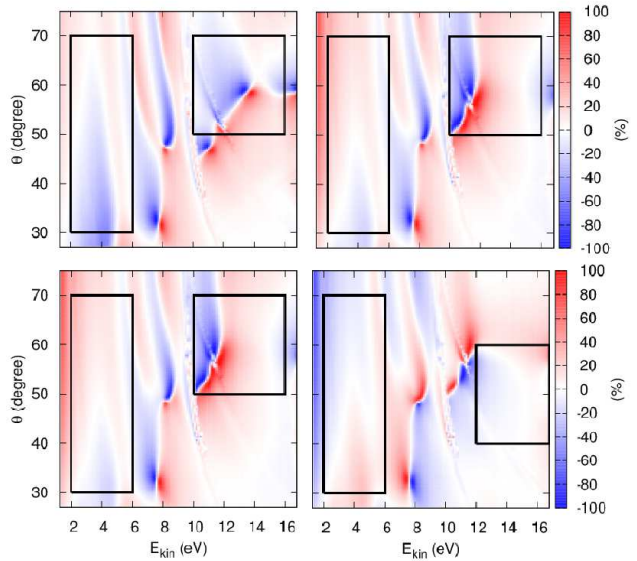}
\caption{
\textbf{top panel:}
Diffraction pattern of the exchange asymmetry ($A_{+}$) for 3 ML Fe/BTO. 
The magnetization of the surface is along the [100] direction (color online).
Left: The polarization of the BTO is directed along the surface normal (P$_{up}$). 
Right: The polarization of the BTO is directed in opposite direction (P$_{down}$).
\textbf{bottom panel:}
Diffraction pattern of the exchange asymmetry ($A_{-}$) for 3 ML Fe/BTO.
The magnetization of the surface is along the [$\overline{1}$00] direction (color online).
Left: The polarization of the BTO is directed along the surface normal (P$_{up}$). 
Right: The polarization of the BTO is directed in opposite direction (P$_{down}$).}
\label{theta_energy_maps_asym_pup_pdown_3ml_magn}
\end{figure}

In Fig. \ref{theta_energy_maps_asym_pup_pdown_3ml_magn} the exchange
asymmetries for 3 ML Fe/BTO for both BTO polarizations are shown
with an [100] magnetization direction of the surface.
For the investigated range of polar angles and kinetic energies two areas (marked by black rectangles)
occur with a large variation of the exchange scattering.

The first area is located at lower kinetic energies of the incident electrons ($\le$ 6 eV)
the second area at higher kinetic energies (12-17 eV).
The different colors in the diffraction patterns indicate
a change of the alignment of electron polarization and
surface magnetization.
Whereas the red color correspond to
a parallel alignment of polarization and
magnetization the blue color correspond to an antiparallel alignment.
As shown in Fig. \ref{theta_energy_maps_asym_pup_pdown_3ml_magn}
changing the BTO polarization from P$_{up}$ to P$_{down}$
changes the reflected spin polarization in the marked areas.
Especially for the area at lower kinetic energy
and higher polar angle, which provides
high reflectivity, a change of the reflected spin
direction due to a change of the BTO polarization
is possible.

In Fig. \ref{theta_energy_maps_asym_pup_pdown_3ml_magn} 
the exchange scattering for both polarization directions of the BTO
for a magnetization in $[\overline{1}00]$ direction is shown in addition.
In comparison to a magnetization direction pointing in the $[100]$ direction
a less pronounced change in the exchange scattering occurs
especially for higher kinetic energies.
This results from spin-orbit scattering caused by the fact that
for a scattering configuration with vanishing spin-orbit interaction
$A_{+}=-A_{-}$ holds \cite{tamura1}, i.e. the exchange asymmetries
for both magnetization directions are indistinguishable.

\begin{figure}[h]
\hspace{-1.cm}
\begin{minipage}[c]{.2\textwidth}
\includegraphics[scale=0.30,angle=270,clip=true,trim=2cm 4.cm 1.cm 7.5cm]{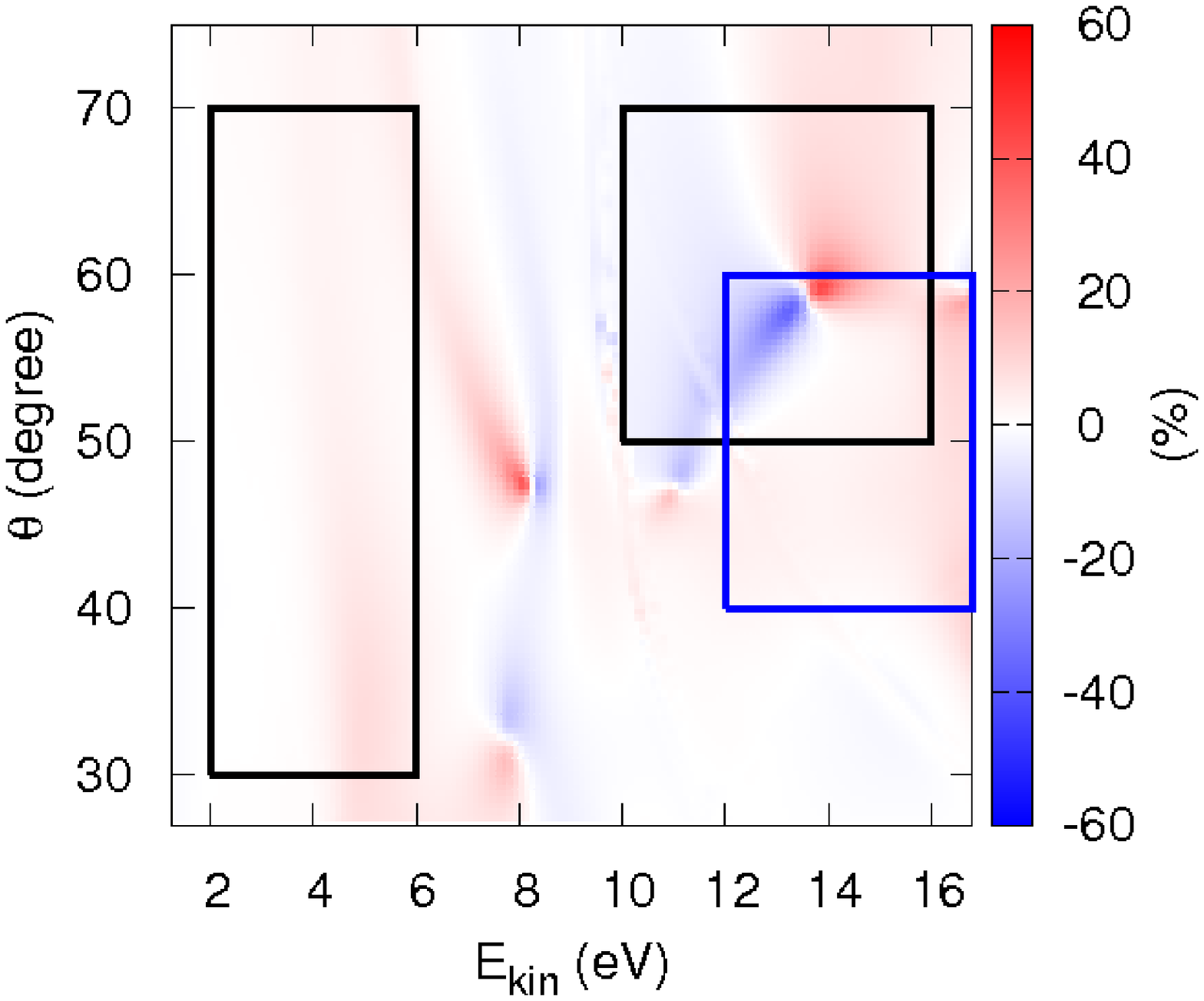}
\end{minipage}
\hspace{0.3cm}
\begin{minipage}[c]{.2\textwidth}
\includegraphics[scale=0.30,angle=270,clip=true,trim=2cm 6.2cm 1.cm 4cm]{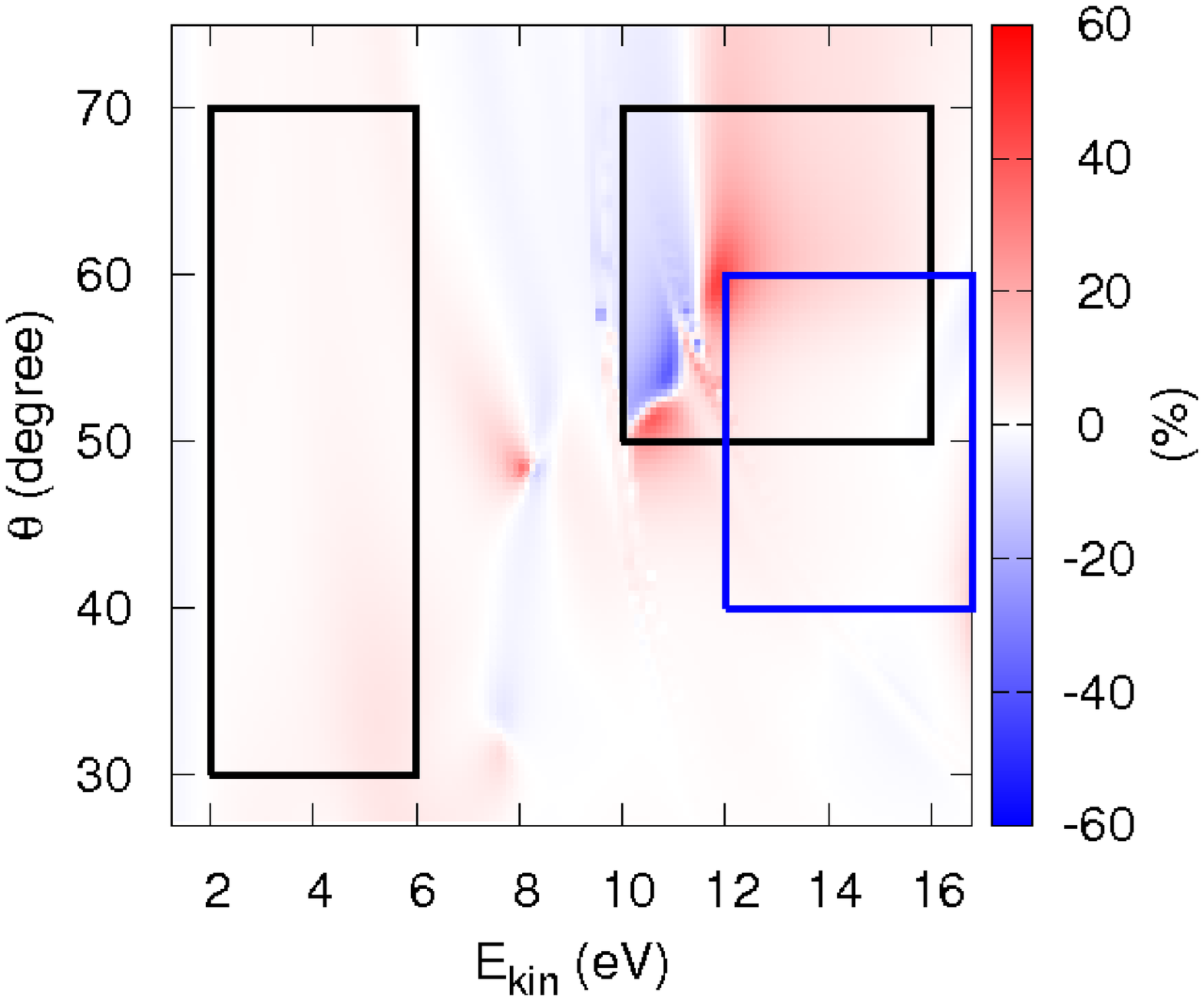}
\end{minipage}
\caption{Diffraction pattern of the spin-orbit asymmetry ($A_{soc}$) for 3 ML Fe/BTO (color online).
Left: The polarization of the BTO is directed along the surface normal (P$_{up}$). 
Right: The polarization of the BTO is directed in opposite direction (P$_{down}$).}
\label{theta_energy_maps_asym_soc_pup_pdown_3ml}
\end{figure}

The spin-orbit induced contribution to the specular scattering of the electrons is shown in Fig.
\ref{theta_energy_maps_asym_soc_pup_pdown_3ml}. Marked by black rectangles
the spin-orbit asymmetry shows a higher change by switching the BTO polarization.
These areas coincide with the rectangles for the exchange asymmetry for
a $[100]$ direction of the magnetization. The blue rectangles correspond to
an exchange asymmetry for the $[\overline{1}00]$ direction. According to the
specified areas of the exchange asymmetry the blue rectangles
show a minor change of the spin-orbit asymmetry and therefore
a minor change in the exchange asymmetry (see Fig. \ref{theta_energy_maps_asym_pup_pdown_3ml_magn}).
This results in a more sensitive reaction of the exchange asymmetry
on a change of the BTO polarization using a $[100]$ orientation of the magnetization.
This is important for the use as
spin polarizing mirror indicating that
an orientation of the surface magnetization
would be preferable along the $[100]$ direction.

\begin{figure}[h]
\hspace{-1.cm}
\begin{minipage}[c]{.2\textwidth}
\includegraphics[scale=0.30,angle=270,clip=true,trim=2cm 4.cm 1.cm 7.5cm]{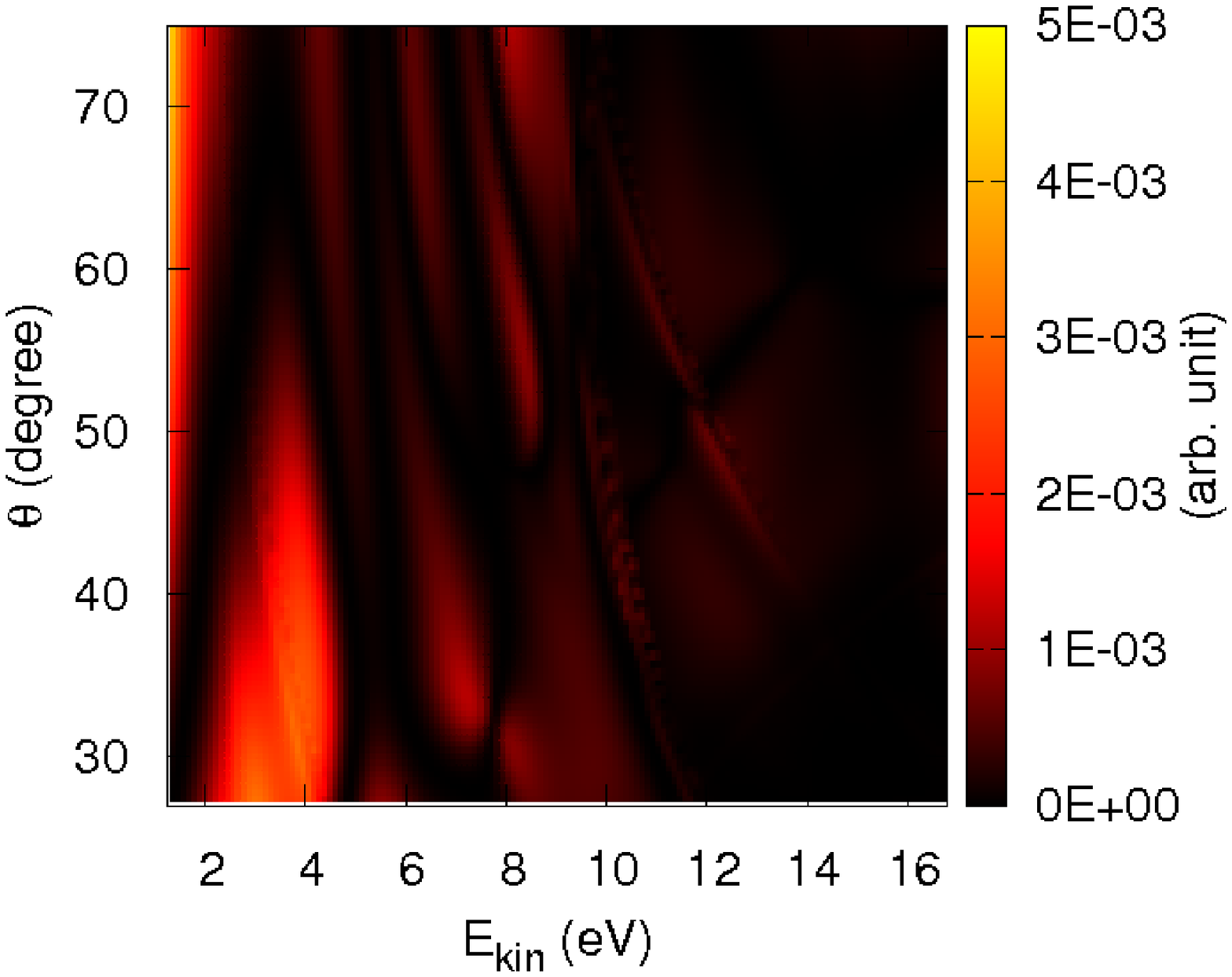}
\end{minipage}
\hspace{0.3cm}
\begin{minipage}[c]{.2\textwidth}
\includegraphics[scale=0.30,angle=270,clip=true,trim=2cm 6.2cm 1.cm 3.8cm]{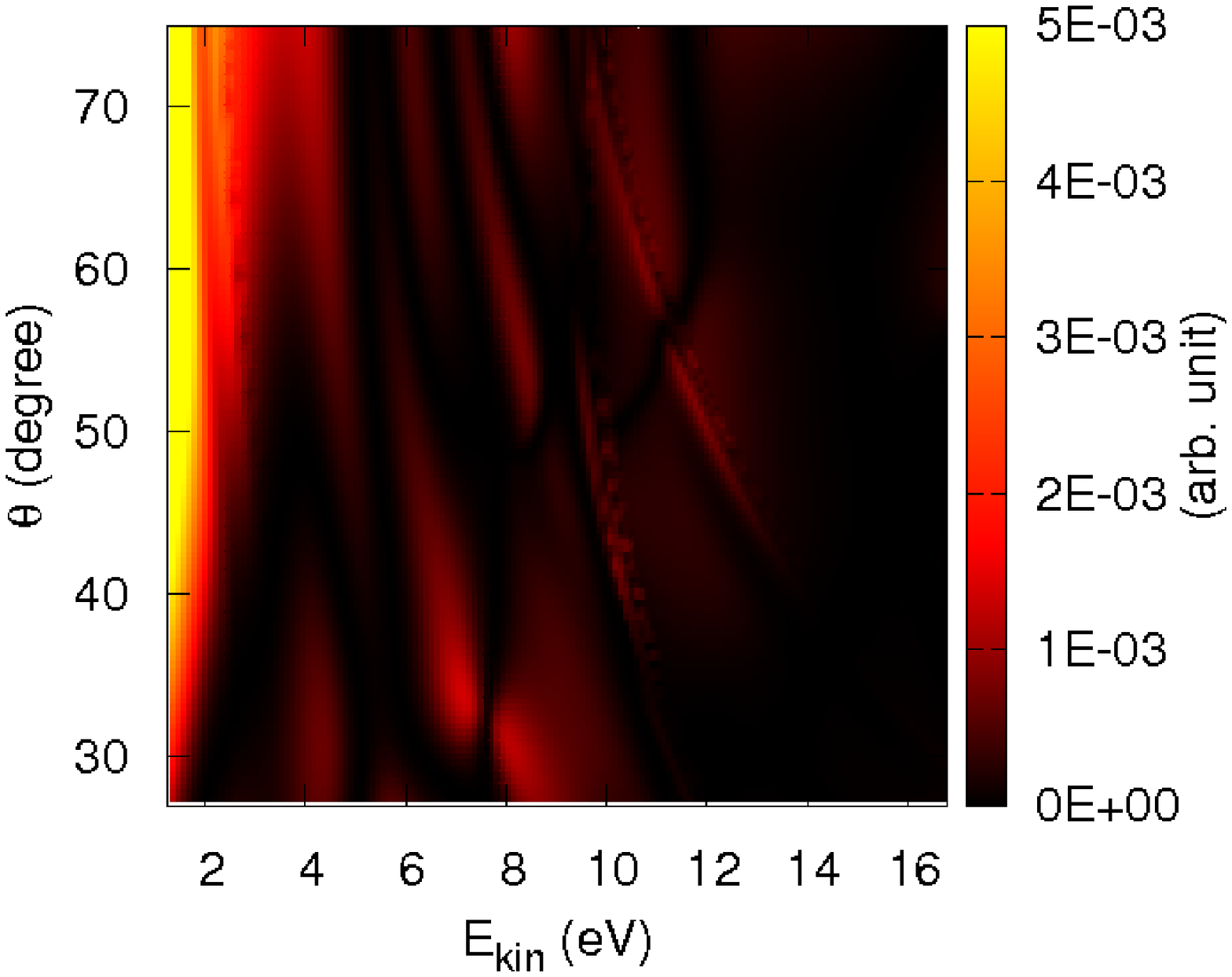}
\end{minipage}
\caption{
Diffraction pattern of the FOM for a [100] orientation
of the magnetization direction for 3 ML Fe/BTO (color online).
Left: The polarization of the BTO is directed along the surface normal (P$_{up}$). 
Right: The polarization of the BTO is directed in opposite direction (P$_{down}$).}
\label{theta_energy_maps_fom_p_pup_pdown_3ml}
\end{figure}

The lowering of the reflectivity affects the FOM via Eq. (\ref{eq:fom}).
The exchange asymmetry enters Eq. (\ref{eq:fom}) to the
power of two and therefore dominantly determines the FOM.
This is the reason for the correspondence of areas
of high exchange scattering and FOM.
In Fig. \ref{theta_energy_maps_fom_p_pup_pdown_3ml} the FOM for both 
BTO polarization directions are shown.

The FOM shows significant changes
for kinetic energies $\le$ 6 eV. Therefore the multiferroic
heterostructure 3 ML Fe/BTO
would have most promising technical applicability for VSPLEED experiments.
With respect to the area of high FOM around 4 eV
for an polarization of the BTO
oriented along the surface normal
the system Fe/BTO could be used as switch for
the electron spin directions. Changing the
BTO polarization the value of the FOM drops
drastically, i.e. no electrons are reflected.

In comparison to investigations done for an
oxygen passivated Fe surface \cite{thiede1,BBM+15}
the reflectivity is in the same
order whereas the FOM is an order of magnitude
lower for the multiferroic heterostructure.
Therefore the spin resolving power is less
for the Fe/BTO system lowering its applicability
as spin-polarizing mirror. Because it is know
that Fe(001)-p(1x1)-O has a longer
lifetime in vacuum and additionally a higher
FOM \cite{BOC99} in comparison to a Fe(001) surface 
it would be interesting to investigate the effect of an
O overlayer on top of Fe/BTO which will be investigated
in a later work.

\subsection{Layer dependence of SPLEED}

In addition we investigated the change of the electron scattering
for thinner Fe layers (1 ML, 2 ML) on top of BTO. The structural properties
correspond to systems studied in previous works \cite{fechner2}.
The results of our calculations
have shown that the most pronounced changes of the exchange
asymmetry by changing the polarization of the BTO
occur for 3 ML Fe/BTO. Therefore the
results for 1 ML and 2 ML will not be shown in detail here.
Nevertheless, the scattering patterns for 1 ML and 2 ML reveal differences between
both BTO polarizations corresponding to 3 ML Fe/BTO.
The differences in the exchange scattering 
between 1 ML, 2 ML and 3 ML Fe/BTO
can be related to the changes of the
in plane spin magnetic moments
changing the polarization of the BTO.
In Table \ref{tab:diff_spin_moments} the changes
of the in plane spin magnetic moments for a
change of the BTO polarization are listed for
the several Fe layers. The largest change
of the spin moments occur for the topmost
Fe layer of 3 ML Fe/BTO. Therefore an electron
impinging on the surface interacting mainly with
the topmost Fe layer would adept the largest change
of the exchange scattering for 3 ML Fe/BTO.
For 1 ML and 2 ML Fe/BTO the changes are smaller
resulting in a less pronounced change of the
exchange scattering.

\begin{table}[H]
\centering
\begin{tabular}{l|c c c c c}
\toprule
& \multicolumn{1}{c}{1 ML} & \multicolumn{1}{c}{2 ML} & \multicolumn{1}{c}{3 ML} \\ 
%\midrule
Fe$\rm_{I}$ &  &  & -0.18 \\ 
 Fe$\rm_{Ti}$+Fe$\rm_{Ba}$ &  & -0.08 & 0.04 \\ 
Fe$\rm_{O}$ & -0.02 & 0.0 & 0.2 \\ 
%\bottomrule
\end{tabular}
\caption{Layer resolved change of the 
in plane spin magnetic moments ($\Delta m_s=m_s(P_{up})-m_s(P_{down})$)
by changing the polarization of BTO.
Units are given in [$\mu_B$].}
\label{tab:diff_spin_moments}
\end{table}

\section{Summary \label{sec_summary}}

Using an ab initio formalism we calculated SPLEED patterns for
1, 2 and 3 ML Fe/BTO. We investigated the change of the reflectivity,
the FOM and both exchange and spin-orbit asymmetry changing the
polarization of BTO. We showed that the 
largest differences of the
diffraction patterns by changing the electric polarization
of BTO results for 3 ML Fe/BTO.
We located two areas in the diffraction patterns
for which a change of the BTO polarization induces a
significant change of the exchange asymmetry.
Therefore a combination of a ferroelectric and a ferromagnetic
material would enable new applications for spin filter purposes,
according to the working areas especially for VSPLEED experiments.
For higher kinetic energies the Fe/BTO system has
a smaller FOM compared to the intensively studied
Fe(001)-p(1x1)-O surface. Therefore investigating the
impact of an additional O layer on top of Fe
will be done in the near future.

\section{Acknowledgement}
We thank the BMBF (05K13WMA), the DFG (FOR 1346),
CENTEM PLUS (LO1402),
CENTEM (CZ.1.05/2.1.00/03.0088) and the COST Action MP 1306 for financial support.

\bibliographystyle{/home/sbo/bibtex/aipnum}
\bibliography{spleed_fe_bto}

\end{document}